\documentclass[12pt,
aps,
 reprint,
 superscriptaddress,
longbibliography
]{revtex4-1}
\usepackage{dcolumn}
\usepackage{bm}
\usepackage%
{graphicx}
\usepackage{xfrac}
\usepackage{extdash}
\usepackage{epsfig}
\usepackage{isomath}
\usepackage{textcomp}
\usepackage{bm}
\usepackage[english]{babel}
\usepackage{todonotes}
\usepackage{color,soul}
\usepackage{times}

\DeclareFontFamily{U}{euc}{}
\DeclareFontShape{U}{euc}{m}{n}{<-6>eurm5<6-8>eurm7<8->eurm10}{}%
\DeclareSymbolFont{AMSc}{U}{euc}{m}{n} 
\DeclareMathSymbol{\umu}{\mathord}{AMSc}{"16}

\newcommand{\ensuretext}[1]{\ensuremath{\text{#1}}}

\newcommand{\unit}[1]{\ensuretext{\textrm{\,}}\ensuremath{\mathrm{#1}}}

\newcommand{\Mum}{\ensuremath{\umu}\ensuremath{\mathrm{m}}}
\newcommand{\mum}{\textrm{\,\ensuremath{\mathrm{\Mum}}}}
\newcommand{\eqref}[1]{(\ref{#1})}
\begin{document}


\title{Observation of ultrafast solid-density plasma dynamics using femtosecond X-ray pulses from a free-electron laser} 

\author{Thomas Kluge}
\affiliation{Helmholtz-Zentrum Dresden-Rossendorf,Bautzner Landstra\ss e 400, 01328, Dresden, Germany}
\email{.kluge@hzdr.de}
\author{Melanie R\"odel}
\affiliation{Helmholtz-Zentrum Dresden-Rossendorf,Bautzner Landstra\ss e 400, 01328, Dresden, Germany}
\affiliation{Technical University Dresden,01069 Dresden, Germany}
\author{Josefine Metzkes}
\affiliation{Helmholtz-Zentrum Dresden-Rossendorf,Bautzner Landstra\ss e 400, 01328, Dresden, Germany}
\author{Alexander Pelka}
\affiliation{Helmholtz-Zentrum Dresden-Rossendorf,Bautzner Landstra\ss e 400, 01328, Dresden, Germany}
\author{Alejandro Laso Garcia}
\affiliation{Helmholtz-Zentrum Dresden-Rossendorf,Bautzner Landstra\ss e 400, 01328, Dresden, Germany}
\author{Irene Prencipe}
\affiliation{Helmholtz-Zentrum Dresden-Rossendorf,Bautzner Landstra\ss e 400, 01328, Dresden, Germany}
\author{Martin Rehwald}
\affiliation{Helmholtz-Zentrum Dresden-Rossendorf,Bautzner Landstra\ss e 400, 01328, Dresden, Germany}
\affiliation{Technical University Dresden,01069 Dresden, Germany}
\author{Motoaki Nakatsutsumi}
\affiliation{European XFEL, Holzkoppel 4, 22869 Schenefeld, Germany}
\author{Emma E. McBride}
\affiliation{European XFEL, Holzkoppel 4, 22869 Schenefeld, Germany}
\affiliation{SLAC National Accelerator Laboratory, 2575 Sand Hill Rd, Menlo Park, CA 94025, USA}
\author{Tommy Sch\"onherr}
\author{Marco Garten}
\affiliation{Helmholtz-Zentrum Dresden-Rossendorf,Bautzner Landstra\ss e 400, 01328, Dresden, Germany}
\affiliation{Technical University Dresden,01069 Dresden, Germany}
\author{Nicholas J. Hartley}
\affiliation{Helmholtz-Zentrum Dresden-Rossendorf,Bautzner Landstra\ss e 400, 01328, Dresden, Germany}
\author{Malte Zacharias}
\affiliation{Helmholtz-Zentrum Dresden-Rossendorf,Bautzner Landstra\ss e 400, 01328, Dresden, Germany}
\affiliation{Technical University Dresden,01069 Dresden, Germany}
\author{Arthur Erbe}
\affiliation{Helmholtz-Zentrum Dresden-Rossendorf,Bautzner Landstra\ss e 400, 01328, Dresden, Germany}
\author{Yordan M. Georgiev}
\affiliation{Helmholtz-Zentrum Dresden-Rossendorf,Bautzner Landstra\ss e 400, 01328, Dresden, Germany}
\affiliation{On leave of absence from Institute of Electronics, Bulgarian Academy of Sciences, Sofia, Bulgaria}
\author{Eric Galtier}
\author{Inhyuk Nam}
\author{Hae Ja Lee}
\author{Siegfried Glenzer}
\affiliation{SLAC National Accelerator Laboratory, 2575 Sand Hill Rd, Menlo Park, CA 94025, USA}
\author{Michael Bussmann}
\affiliation{Helmholtz-Zentrum Dresden-Rossendorf,Bautzner Landstra\ss e 400, 01328, Dresden, Germany}
\author{Christian Gutt}
\affiliation{Universit\"at Siegen, Adolf-Reichwein-Stra\ss e 2, 57068 Siegen, Germany}
\author{Karl Zeil}
\affiliation{Helmholtz-Zentrum Dresden-Rossendorf,Bautzner Landstra\ss e 400, 01328, Dresden, Germany}
\author{Christian R\"odel}
\affiliation{Friedrich-Schiller-Universit\"at, F\"urstengraben 1, 07743 Jena, Germany}
\author{Uwe H\"ubner}
\affiliation{Leibniz Institute of Photonic Technology, Albert-Einstein-Stra\ss e 9, 07745 Jena, Germany}
\author{Ulrich Schramm}
\author{Thomas E. Cowan}
\affiliation{Helmholtz-Zentrum Dresden-Rossendorf,Bautzner Landstra\ss e 400, 01328, Dresden, Germany}

\date{}

\begin{abstract}
The complex physics of the interaction between short pulse high intensity lasers and solids is so far hardly accessible by experiments. 
As a result of missing experimental capabilities to probe the complex electron dynamics and competing instabilities, this impedes the development of compact laser-based next generation secondary radiation sources, e.g. for tumor therapy\cite{Bulanov2002,ledingham2007laser}, laboratory-astrophysics\cite{Remington1999,Bulanov2015a}, and fusion\cite{Tabak2014}. 
At present, the fundamental plasma dynamics that occur at the nanometer and femtosecond scales during the laser-solid interaction can only be elucidated by simulations. 
Here we show experimentally that small angle X-ray scattering of femtosecond X-ray free-electron laser pulses facilitates new capabilities for direct \emph{in-situ} characterization of intense short-pulse laser plasma interaction at solid density that allows simultaneous nanometer spatial and femtosecond temporal resolution, directly verifying numerical simulations of the electron density dynamics during the short pulse high intensity laser irradiation of a solid density target. 
For laser-driven grating targets, we measure the solid density plasma expansion and observe the generation of a transient grating structure in front of the pre-inscribed grating, due to plasma expansion, which is an hitherto unknown effect. 
We expect that our results will pave the way for novel time-resolved studies, guiding the development of future laser-driven particle and photon sources from solid targets. 
\end{abstract}
\maketitle 

The solid density plasmas created in the interaction of an ultra-short, ultra-high intensity (UHI) laser pulse with a solid target are a source of femtosecond, high-charge  electron\cite{Thevenet2015} and ion bunches\cite{Hatchett-ElectronPhotonAndIons,fuchs,Schwoerer-MonoenergeticProtons,Macchi2013b}, extreme ultraviolet (XUV) radiation\cite{Dromey-HighHarmonic,Roedel2012,Monchoce2014}, and neutrons\cite{Roth2013}, making them promising candidates for future particle accelerators or radiation sources. 
Until now a fundamental impediment of the ongoing research of UHI laser-solid interactions has been the limited experimental capability of diagnosing the basic processes during the laser interaction on the relevant scales that range from sub-femtosecond to hundreds of femtoseconds and from few nanometers to few hundred nanometers. 
Some of the most important physical processes are, for example, the generation of plasma oscillations~\cite{Sherlock2014} and plasma waves~\cite{Kluge2015}, electron transport and plasma heating\cite{Leblanc2014,kluge2015current}, instability development~\cite{Sentoku2000,Borghesi2011,Palmer2012,Metzkes2014,Kluge2015,Sgattoni2015,Gode2017}, and the generation of strong magnetic fields~\cite{Leblanc2014}. 
A fundamental process is the expansion of the irradiated plasma into vacuum~\cite{Mora2003,mora:056401,schreiber2006analytical} during the laser interaction, governing the surface dynamics and laser absorption both prior to and during the laser main pulse. 
\begin{figure*}
\centering
  \begin{minipage}[b]{0.55\textwidth}
      \includegraphics[width=\linewidth]{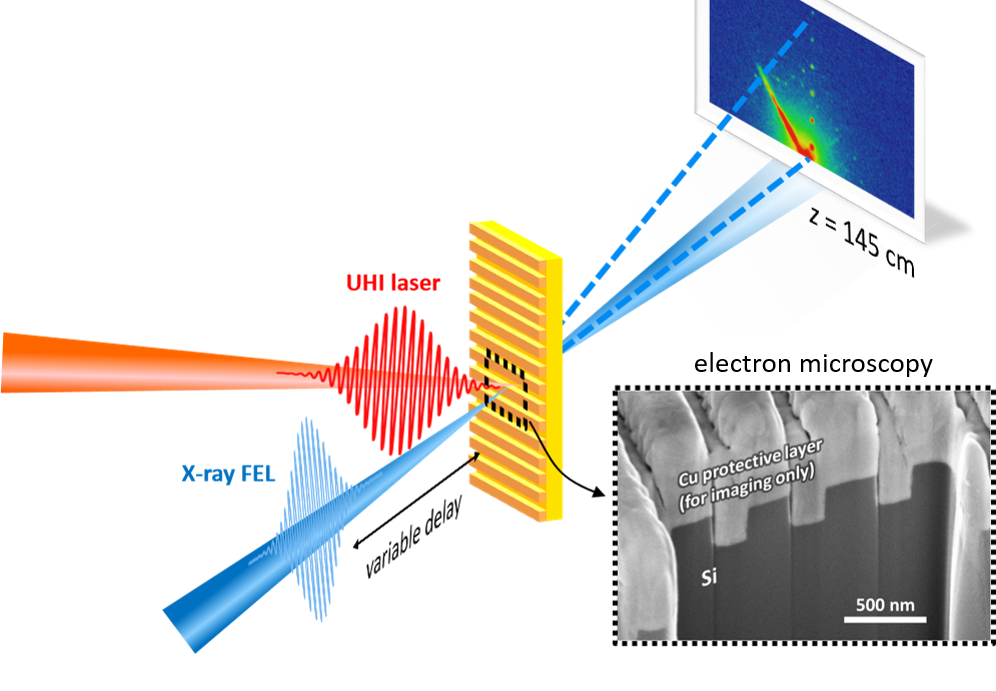}
%
%
  \end{minipage}
  \begin{minipage}[b]{0.06\textwidth}
  ~
  \end{minipage}
  \begin{minipage}[b]{0.37\textwidth}
      \includegraphics[width=\linewidth]{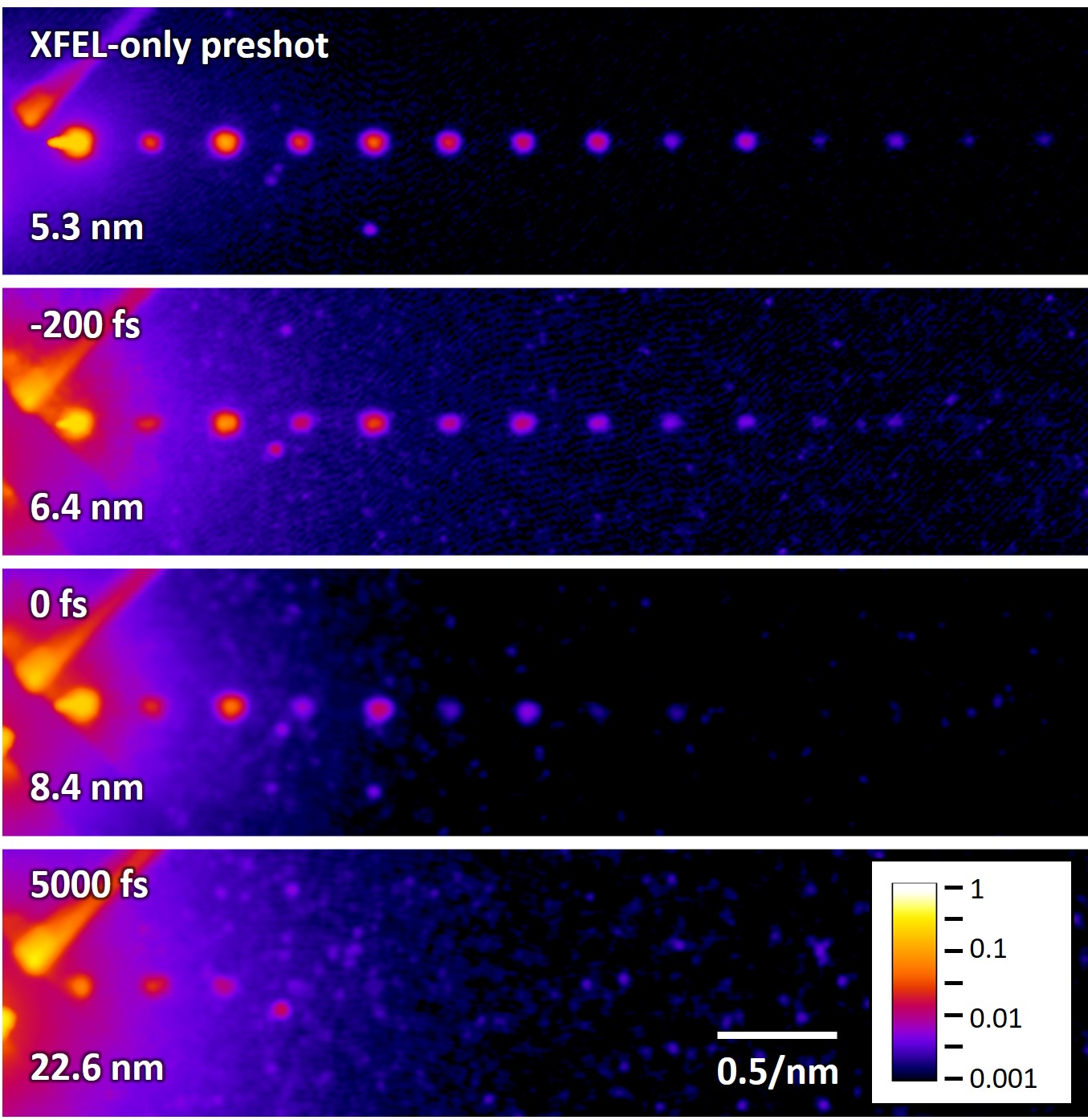}
  \end{minipage}
\caption{\textbf{Schematic of the SAXS experimental setup.} \textbf{left} A near infrared laser pulse ($5\cdot 10^{17}\unit{W/cm}^2$, duration $80\unit{fs}$) is focused to a spot of $\approx 16\unit{\mum}\times 30\unit{\mum}$ (FWHM), irradiating a silicon grating target under $45^\circ$ parallel to the grating ridges and leading to an ultrahigh intensity (UHI) laser plasma interaction. The XFEL pulses ($40\unit{fs}$, $5-10\unit{\mum}$ FWHM spot) probe the laser plasma dynamics under $90^\circ$. The inset shows an electron scanning microscopy image of a representative target sample with grating period $500\unit{nm}$ (targets were covered with Cu only for imaging). 
\textbf{right} Examples of normalized scattering patterns of a grating with period $g=300\unit{nm}$  for different XFEL delays (rotated by $90^\circ$).
}
\label{fig:setup}
\end{figure*}

For each application a correspondingly tailored surface structure can enhance laser absorption and interaction, electron acceleration, and hence all subsequent processes. 
In fact, it has been shown that a preplasma density gradient, e.g. generated by laser intensity prior to the main pulse
, strongly affects absorption\cite{Ping2008} and the generation of secondary radiation such as proton acceleration\cite{kaluza2004ilp} and XUV radiation\cite{Kahaly2013}. 
Grating targets have even shown nearly complete absorption\cite{Kahaly2008}, resulting in enhanced ion acceleration\cite{Ceccotti2013} and XUV generation\cite{Cerchez2013}.

%
%
Measuring, controlling and eventually optimizing collective particle motion excited by, and interleaved with, the intense laser light is the key to pushing the development of laser based radiation sources further. 
Conventionally, these dynamic effects can only be accessed by numerical simulations, indirect experimental methods (e.g. particle/radiation emission from the bulk) or optical methods\cite{Geindre1994,Mondal2010}, which are however all limited in either spatial or temporal resolution or to the optically transparent low density regions. 

The present work aims at making the few femtosecond, few nanometer regime directly accessible at solid density by employing small angle X-ray scattering (SAXS) using X-ray free electron laser (XFEL) pulses\cite{O.Glatter1982,Kluge2014,Gorkhover2016}. 
With the advent of XFELs there have been many approaches to use them as a probe for UHI laser-driven solid density plasmas. 
X-ray beams can penetrate solid density plasmas and hence can resolve the dynamics at high densities throughout the whole target. 
Taking advantage of the unique properties of XFEL pulses, e.g. their short duration, high number of photons per bunch, high transverse coherence, monochromaticity and polarization, the XFEL can widely expand the capabilities of established diagnostic tools. 
%
%
While X-ray diffraction\cite{Gauthier2014} typically provides a resolution on the order of angstroms, direct imaging methods such as phase contrast imaging\cite{Schropp2015} are limited to the X-ray spot size of a few hundred nanometers. 
SAXS can bridge the gap with a resolution between the nanometer and few micron scale.

Here, we present the first measurement of the spatial density distribution of intense short-pulse laser-driven solid density plasma to the nanometer level. 
We use SAXS of XFEL pulses which 
allow for a  single shot direct electron density measurement, and study \textit{in situ} the plasma expansion during the laser irradiation. 
Employing the pump-probe technique, we can follow the expansion of the plasma at solid density in real time and compare directly to simulations. \\


%
\subsection*{Results}
\begin{figure*}
\centering
  \includegraphics[width=\linewidth]{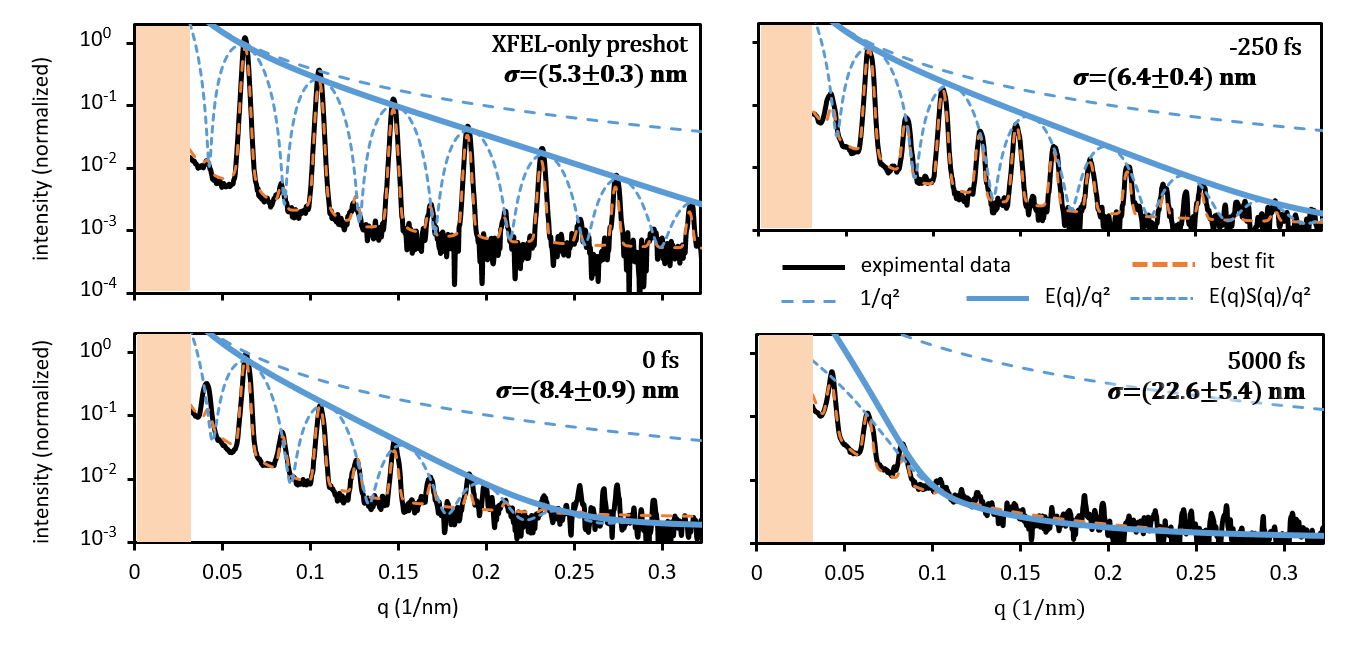}
\caption{\textbf{Lineouts and fitting:} Cuts through the horizontal direction through the scattering peaks of the SAXS pattern for the four shots shown in the right of Fig.~\ref{fig:setup} (black lines) and fitting results. From the fits (dashed orange lines) the individual contributing terms of Eqn.~\eqref{eqn:scatteringI} are shown: $1/q^2$ corresponds to scattering from a perfectly sharp step (blue long dashed lines); $E(q)/q^2$ from a smooth edge with expansion parameter $\sigma$ (blue solid lines); and $S(q)E(q)/q^2$ from a single slit with smooth edges (blue short dashed lines). Orange area marks the region with over-saturated pixels and was not used for fitting. 
}
  \label{fig:results1}
\end{figure*}
\begin{figure}
\centering
  \includegraphics[width=\linewidth]{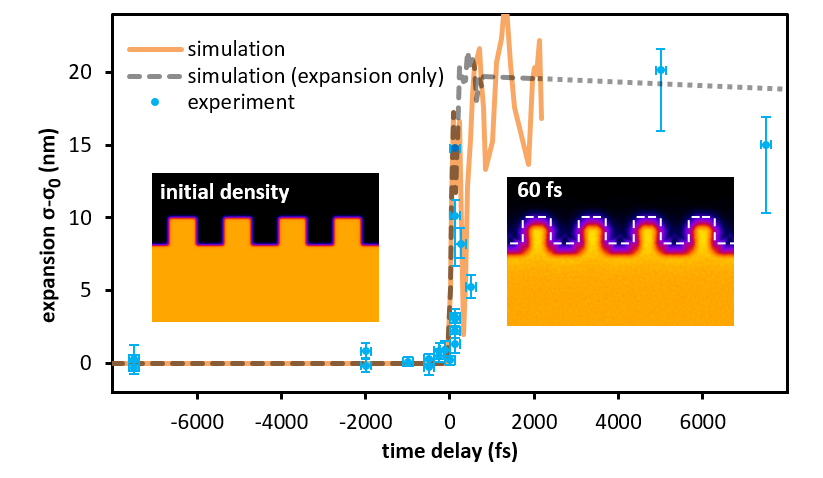}
\caption{\textbf{Surface expansion into vacuum:} Expansion parameter $\sigma-\sigma_0$ as a function of XFEL delay extracted from the scattering patterns (blue circles). Orange solid (gray dashed) line shows $\sigma$ extracted from the simulation by fitting a grating with error-function shaped edges to the density integrated through the target, including (excluding) the plasma jets emitted between the grating ridges (dotted line is the extrapolation of the dashed line beyond the simulated delay time of $2\unit{ps}$). Excluding the jets, $\sigma$ resembles the expansion of the ridges, while including them $\sigma$ corresponds to the shape of the plasma jets when their density dominates over that of the remaining pre-inscribed ridges at delays larger than $\approx 270\unit{fs}$ (see main text for details). The insets show the plasma density from simulation (color scale same as in Fig.~\ref{fig:double}). }
  \label{fig:scan}
\end{figure}

The experiments were performed at the Matter in Extreme Conditions (MEC) endstation of LCLS at SLAC\cite{Nagler2015,Fletcher2015} with the Ti:Sa near infrared short-pulse laser (see setup in Fig.~\ref{fig:setup}). 
Synchronization with the LCLS XFEL beam was achieved to an accuracy of $120\unit{fs}$, measured by a spatially encoding autocorrelator~\cite{Bionta2011,Harmand2013}. 
Our samples were $2\unit{\mum}$ thick silicon membranes with an inscribed surface grating. 
Additionally to being a fundamentally interesting target for optimizing the laser-solid interaction applications, the advantage of using tailored grating surface membranes is the large increase in scattering signal compared to a flat surface. 
We can now follow small variations in the scattering signal to detect few nanometer structural changes, induced by the laser irradiation, with a high spatial resolution reaching down to a nm and below. \\
In the following, we present a series of shots on surface gratings with periods $g$ ranging from $150\unit{nm}$ to $500\unit{nm}$ where the ridge widths $b$ were approximately half of the grating period and the depth $100\unit{nm}$. 
The rectangular ridges and the membrane itself were flat to the level of $\sigma_0=(5.4 \pm 0.7)\unit{nm}$ which was measured for each target by two consecutive XFEL-only SAXS measurements at reduced intensity that did not damage the target\footnote{Only one target showed a significantly larger initial ridge sharpness of $\sigma_0=11.3\unit{nm}$, marked by a dark dot in Fig.~\ref{fig:scan}}. 
Those grating dimensions, besides being in the relevant range for absorption enhancing surface structures, are also in the range of the spatial dimensions of instabilities in relativistically driven plasmas and are therefore also interesting on a technical level for future applications of SAXS. 

In the regime relevant for this work, the dominant source of small angle X-ray scattering is coherent elastic Thomson scattering from electrons. 
In the SAXS geometry the scattering pattern is obtained in Born approximation by the absolute square of the exit wave Fourier transform\cite{O.Glatter1982}. 
With an appropriate model for the density in real space, the gradient of the expanded plasma and other spatial features can be characterized by fitting the respective correlation function to the scattering pattern in reciprocal space\cite{Kluge2017}. 

Typical scattering patterns are shown in Fig.~\ref{fig:results1} for three different XFEL-laser delay times. 
The scattering signal covers the whole field of view of the detector, with decreasing intensity at larger scattering vector values $q$. 
With increasing delay time we observe a decrease of the scattering signal at large $q$-values. 
We interpret this as an exponential roll-off due to the expansion of the plasma bulk, in analogy to the Debye Waller factor.  \\
We model the plasma line density along the grating direction as the convolution of a grating delta comb, a single slit (ridge), and the ridge edge shape described by a Gaussian, $\left(\sfrac{N_b}{b}\right)\exp(-x^2/2\sigma^2)$. 
The density along a ridge edge is then given by an error function, $\tilde{n}_e\equiv\left(\sfrac{N_b}{2b}\right) \left[ \mathrm{erf}\left(\sfrac{x}{\sqrt{2}\sigma}\right)+1\right]$, where $N_b$ is the total number of electrons in a ridge in the illuminated area. 
The expansion parameter $\sigma$ defines the smoothness of the grating ridge and is therefore a measure of the plasma expansion scale length. \\
The scattered intensity in the distance of a unit length in Born approximation is derived from the product of the respective squares of the normalized Fourier transforms of the grating, $G(q)$, single slit, $S(q)$\cite{Gerthsen1993}, and error function, $E(q)$\cite{Geller1969}, 
\begin{equation}
I\left(q\right)=I_0 r_e^2 \left(2N\frac{N_b}{b}\right)^2 \frac{1}{q^2} G\left(q\right)\cdot S\left(q\right)\cdot E\left(q\right)
\label{eqn:scatteringI}
\end{equation}
with 
\begin{eqnarray}
    \label{eqn:grating}
	G\left(q\right) &=& \left[\frac{\sin{\left(N q g/2\right)}}{N \sin{\left(q g/2\right)}}\right]^2,\\
    \label{eqn:single}
	S\left(q\right) &=& \sin^2{\left(\frac{q b}{2}\right)}, \\
    \label{eqn:waller}
    E\left(q\right) &=& e^{-q^2\sigma^2}
\end{eqnarray}
($I_0$: XFEL intensity, $r_e$: classical electron radius, $N$: number of ridges in the illuminated area).  

In Fig.~\ref{fig:results1} we illustrate the quantitative determination of the time dependent plasma expansion by describing the data using Eqn.~\eqref{eqn:scatteringI} as an estimator. 
The single slit scattering envelope $S\left(q\right)$ from Eqn.~\eqref{eqn:single} causes some peaks described by the grating spatial frequency harmonics in $G\left(q\right)$ to be suppressed due to the phase matching of $S\left(q\right)$ and $G\left(q\right)$ related to the $b/g$ ratio. 
Finally, the edge shape envelope $E(q)$ from Eqn.~\eqref{eqn:waller} yields the parameter $\sigma$ describing the bulk expansion. 
The full fitting procedure is described in the Methods supplementary. 
\begin{figure}
%
\centering
  \includegraphics[width=\linewidth]{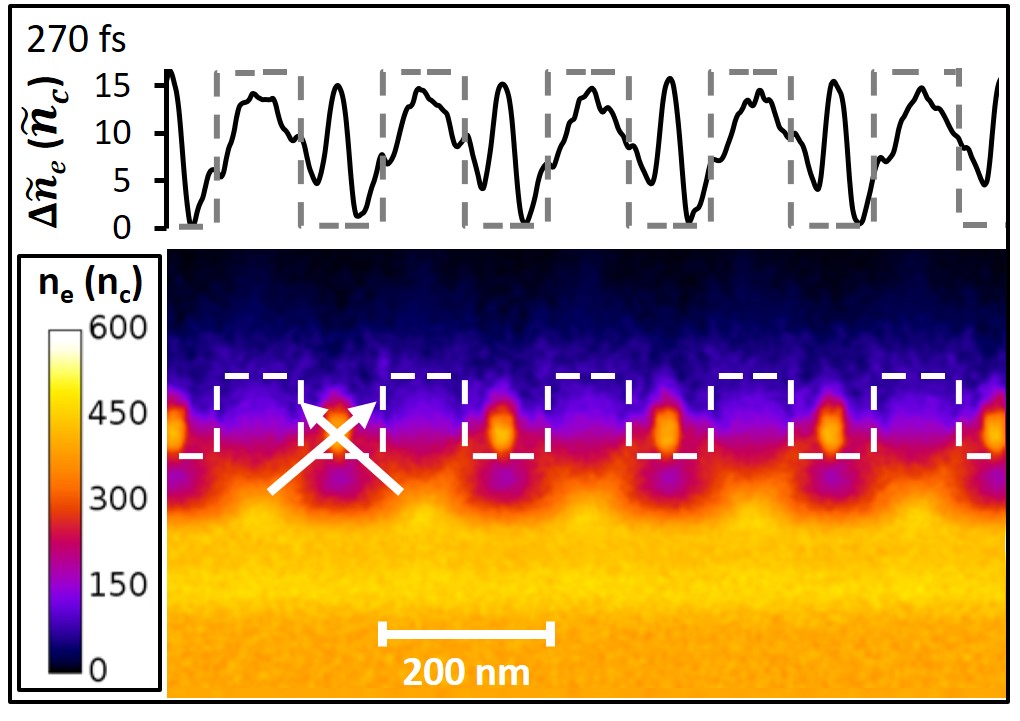}
  \includegraphics[width=\linewidth]{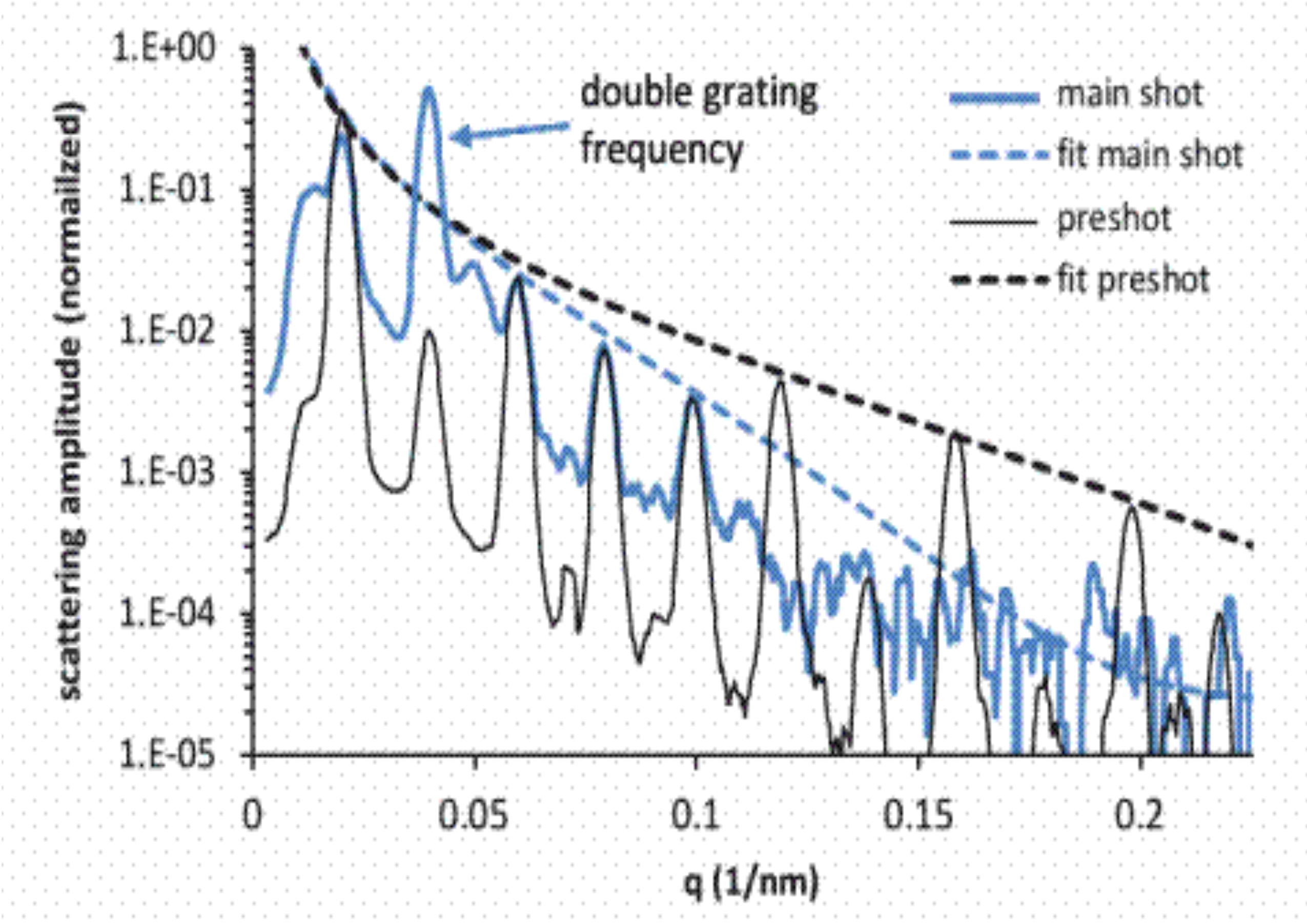}
\caption{\textbf{Transient plasma grating:} \textbf{Top} PIC simulated electron densities suggest the emission of plasma jets (white arrows) from the sharp edges at the surface due to large local fields. Integrating along the vertical direction (top panel), we find the grating frequency  doubled. This repeats periodically whenever the jets intersect. \textbf{Bottom} Cuts through the experimental scattering pattern for the XFEL-only preshot (black line) and main shot (delay 6 ps, blue line), normalized to the same respective fitted amplitude at $q=0$. The dashed lines illustrate $E(q)$ of the best respective fit, excluding the second order scattering feature for the main shot. Its height is much larger than could be explained by scattering off the cold grating structure alone, indicating that the expansion could have effectively formed a double frequency grating. $n_c=1.7\cdot10^{21}\unit{cm}^{-3}$ is the critical electron density. }
%
  \label{fig:double}
\end{figure}

The fitting results for all of our full-intensity laser shots are summarized in Fig.~\ref{fig:scan}, together with particle-in-cell simulation results (orange line, see Methods). 
When the laser pulse hits the target, a fast expansion of the grating ridges is observed. 
The agreement between the measurement and the simulation is excellent: both the fast expansion during the femtosecond optical laser pulse irradiation and the stagnation thereafter are well reproduced. 
We note that despite the laser contrast showing significant prepulses approximately $50-40\unit{ps}$ prior to the main pulse (see Methods for details on the temporal contrast) an expansion of the target bulk prior to the optical laser main pulse was not observed, on the scale of the nanometer to tens of nanometers sensitivity of our setup -- proving ultra.short XFEL diffraction before destruction. \\
We can extract the expansion velocity by linear interpolation during the expansion phase around $t=0$. 
With the timing accuracy limited by the XFEL delay jitter, we can estimate only a lower limit for the expansion velocity of approx. $0.04\unit{nm/fs}$ which compares to $0.19\unit{nm/fs}$ measured in the simulation. 
From the simulation we find that after approximately $100\unit{fs}$ the expansion has progressed so far that the rising and falling edges of the ridges intersect and penetrate each other. 
We infer from the simulation that the ridges eventually even disappear after approximately $200\unit{fs}$ and only density modulations in the compressed density layer remain (see Fig.~\ref{fig:double} top). 
Their edge shape quickly stagnates at an error function width of approximately $25\unit{nm}$ as the driving charge separation neutralizes after the optical laser pulse has passed, $\sigma$ then remains almost constant over the remaining $1.5\unit{ps}$ of the simulation, and likely it does also beyond. 
This is in agreement with the experiment reasonably well, confirming that the compressed front surface layer indeed remains intact over the time course of a few picoseconds. \\

We observe an interesting effect in the simulations of the expanding plasma grating that has not been described so far. 
While two neighboring grating ridges expand and form plasma jets, they intersect so that an additional plasma grating is generated at the intermediate position, see Fig~\ref{fig:double} and the supplementary movie at [URL will be inserted by publisher]. 
In the simulation this happens for the first time around $270\unit{fs}$ after the main pulse, and repeats periodically after approximately a picosecond. 
When the jets cross each other, the density modulation can get larger than that of the expanding ridge remainders, then dominating the measurements of the expansion parameter. 
This leads to the oscillating structure in the expansion parameter extracted from the simulation (Fig.~\ref{fig:scan}, solid orange line). 
For a short time of a few ten femtoseconds the density of the transient grating can match the pre-inscribed grating. 
In the projection integrated along the XFEL both together then form a modulation of twice the initial grating frequency (see Fig.~\ref{fig:double} top). \\ 
Since the SAXS signal is sensitive to the spatial frequencies, we can study this novel effect by analyzing the heights and extinction of the individual peaks in the scattering patterns. 
For the ideal grating, with $b/g\approx 1/2$, $S\left( q \right)$ minimizes near the position of the even harmonics of the grating spatial frequency (compare dashed line in Fig.~\ref{fig:results1}). 
The fabricated gratings slightly deviate from the ideal $b/g$ ratio so that the exact peak heights vary between individual samples and also even harmonics may appear, for $b/g>1/2$ the first minimum of $S(q)$ shifts to larger $q$ and vice versa. 
This means that upon optical laser irradiation, when the grating ridge expands, the second harmonic peak can increase or decrease relative to the first peak, depending on the initial position of the single slit minimum. 
The maximum height of the second peak is however limited to the height of the envelope (Eqn.~\eqref{eqn:scatteringI} with only the exponential term $E(q)$, ignoring $G(q)$ and $S(q)$) as can be seen exemplary in Fig.~\ref{fig:results1}. \\
However, in one shot we observe a peculiar excess scattering intensity at the position of the second harmonic peak (see Fig~4 bottom).  
The only explanation for this is the existence of a structure of twice the pre-inscribed grating frequency in the target density projected along the XFEL direction -- lending evidence that the SAXS method is sensitive also to such detailed dynamics and could be used to optimize the generation of transient nano-grating structures. 
Especially in the context of other mechanisms leading to transient gratings this may become relavant, considering e.g. the power distribution in holraums\cite{Glenzer2010} or harmonic generation\cite{Monchoce2014} where SAXS could be used for sub-laser-cycle diagnostics of the laser-plasma interaction~\cite{Kluge2017}. \\
In the present case the double grating exists only for a few 10s of femtoseconds, acting as a fast switch for the X-ray beam in the direction of the second maximum. 
This explains why we observe excess intensity only in one shot: The timing of the XFEL probe arrival must match the formation of the transient grating quite accurately, so in the majority of shots we would likely miss it. 

\subsection*{Discussion and outlook}
In summary, we presented the first quantitative measurement of plasma expansion at solid density with nanometer, few femtosecond resolution, with a near relativistic high intensity laser. 
It is well-known that the critical density surface is expanded by a significant fraction of the laser wavelength, but how much the solid density surface expands was not well known. 
We have experimentally verified that the surface expands but besides this stays intact during the interaction and a few picoseconds beyond (for the intensity and pulse duration regime we examined). 
While information of the critical density surface is important for laser absorption, information of the solid density dynamics is crucial for understanding complex processes e.g. laser-generated particle transport, bulk heating and instability development, as well as production of static electric and magnetic fields that are generated strongly at the steep density jump (resistivity jump). 

The concept of pre-inscribed gratings serving as a reference and providing strong signal at large $q$ from a well defined spatial region was the key for the high spatial sensitivity. 
Placing the gratings at the rear surface would immediately allow us to characterize the TNSA process\cite{PhysRevLett.84.4108,wilks} where the solid density plasma expansion has never previously been measured on the relevant scales of few nanometers and femtoseconds. 
The grating method could further be extended for example to buried gratings for studying isochoric or buried layer heating\cite{LGHuang,Huang2016}, gratings of a specific material with bound-bound resonances at the XFEL energy for resonant scattering\cite{Kluge2016}, or more complex 2D structures -- e.g. to encode spatial markers into the $q$ space. 
For example, a thin layer of a high density material within a low density material could serve as a spatial limiter for density dependent features such as plasma oscillations, and could help mitigating the problem of plasma motion blur or other volume effects. \\
At higher laser intensities, where plasma instabilities and relativistic effects are expected to lead to spatial modulations at similar scales, SAXS measurements might help to optimize future laser-based particle and radiation sources and therefore will pave the way for novel applications. 

\subsection*{Methods}
\textbf{Optical laser}
\begin{figure}
\centering
  \includegraphics[width=\linewidth]{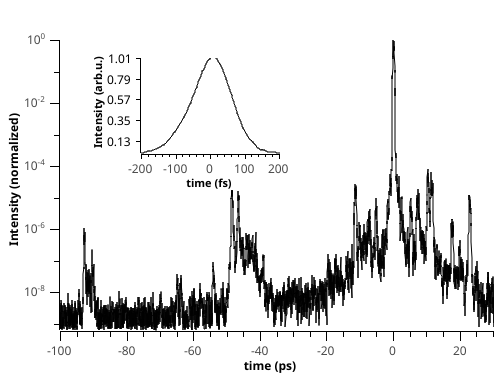}
\caption{\textbf{Laser contrast and pulse duration.} The laser contrast base level was measured by a cross-correlator and is better than $10^{-8}$. This suppresses the laser intensity below the ionization threshold of Si until the prepulse at $-50\unit{ps}$. The main laser pulse temporal shape is shown in the inset by the autocorrelation trace. }
  \label{fig:laser}
%
%
\end{figure}
The optical laser used for target excitation, i.e. plasma generation, is the Titanium:Sapphire-based high power laser system at the MEC end station at LCLS. 
The laser system provides ultra-short pulses ($\tau$ = 80 fs) at a central wavelength of 800 nm, the pulses containing an energy of 1\,J before the compressor and 460\,mJ on target. 
The laser pulse was focused by an off-axis parabolic mirror, the focal spot size ($30\unit{\mum}\times 16\unit{\mum}$ FWHM) was measured as the average of 26 single shot measurements (systematic and statistical error below 2\%).
Within the FWHM of the spot, 22\% of the laser pulse energy are contained, leading to an optical laser intensity (average over FWHM area) of $4 \cdot 10^{17}\unit{W/cm}^2$ which for Gaussian shape corresponds to a peak intensity $5\cdot 10^{17}\unit{W/cm}^2$. 
The intensity on target taking into account the angle of incidence of $45\deg$ is a factor of $\sqrt{1/2}$ smaller. 
The temporal intensity contrast of the laser pulse - i.e. the ratio of the background laser intensity normalized to the peak intensity of the ultrashort pulse - was characterized with a third-order cross-corelator~\cite{Oksenhendler2017}, with a temporal resolution of approx. $100\unit{fs}$, Fig.~\ref{fig:laser}. The detection threshold lies at $\approx$ 2 $\cdot$ 10$^{-9}$. 
The prepulse at $-48\unit{ps}$ is real according to the measurement whereas other pulses occurring at negative delay (i.e. before the main pulse) could be artefacts from the measurement itself. As was shown in Fig.~\ref{fig:scan}, the prepulses do not lead to a measurable expansion of the sample. However, this does not rule out the possibility of formation of a low density preplasma, which however would have to have either a density small enough to not be detected, or a scale length large enough that the exponential roll-off occurs well before the first scattering order. 

\textbf{XFEL} The LCLS beam was focused with compound refractive lenses into the MEC experimental area to a spot size between 5 and $10\unit{\mum}$. It was attenuated by various absorbers to ensure the scattering signal was within the dynamic range of the PIXIS XF 2048B camera. The direct XFEL beam transmitted through the target was blocked by a beam-block wire in front of the vacuum exit window.  We recorded two XFEL preshots before each optical laser shot-- the first to qualify the undriven target structure and the second to prove that the XFEL at the selected intensity does not damage the sample. 

\textbf{Synchronization} A crucial point of the experimental setup and preparation is the spatial and temporal overlap of the optical and XFEL pulse as well as the target at the interaction point.
The interaction point is fixed in space via two independent high resolution alignment imaging systems ($\approx$ 0.2 $\mu$m/px resolution at a field of view of 200 $\times$ 360 $\mu$m$^2$ and 290 $\times$ 390 $\mu$m$^2$, respectively).
For the spatial alignment of the XFEL pulse, a target ($500\mu$m thick Si wafer) is placed in the interaction plane and a burst of XFEL pulses is used to create an $\approx$ 30 $\mu$m diameter damage (hole) in the target, which can be imaged with the alignment imaging systems. This defines the interaction point in the interaction plane.
The optical pulse as well as the target are then aligned to this interaction point using the alignment imaging systems. The overlap of the XFEL pulse with the grating structure was independently confirmed in the non-destructive preshots taken before every combined irradiation of the target by the optical laser and XFEL.\\
The temporal overlap of the optical and XFEL pulse at the interaction point is measured with a $20\unit{\mum}$ thick YAG plate. 
The XFEL pumps the crystal in target position by generating free-carriers in the YAG and therefore changing its refractive index. 
This change in refractive index can be probed by the optical laser to determine the synchronization point. 
This setup is comparable to the spatial encoding \textit{time tool} presented in~\cite{Bionta2011,Harmand2013}. 
The same principle is applied in the time tool installed in the XFEL beam path upstream of the interaction point in order to monitor the shot-to-shot timing jitter between optical and XFEL pulse. Here, the  the ultrafast free-carrier generation in a Si$_3$N$_4$ slab is probed by a pulse split from the main optical pulse. 
The \textit{time tool} measured a timing jitter for the relative arrival time between optical and XFEL pulse $119\unit{fs}$ which determines the delay time error bars in Fig.~\ref{fig:scan}.

\textbf{Small Angle X-ray Scattering}
Depending on the X-ray photon energy and the UHI laser intensity, and therefore ionization degree and opacity of the plasma, the scattering of keV X-ray photons is dominated by elastic or inelastic XRTS or photo-ionization process\cite{Kluge2016}. 
Here we assume that the XFEL photon energy is far enough from bound-bound transition energies, so that excitations and ionization of inner electronic shells of atoms or ions by the XFEL can be ignored. 
For small scattering angles the signal is then dominated by elastic Thomson scattering. \\ 
In the SAXS geometry the scattering image is obtained in Born approximation simply by absolute square of the exit wave Fourier transform\cite{O.Glatter1982}, the latter being given by product of the plasma electron density integrated along the X-ray propagation direction and the Gaussian-like illumination. 
The respective spatial and temporal resolution are fundamentally limited only by the X-ray wavelength and pulse duration -- enabling nanometer and few femtosecond resolution. 
Of course, for a real experiment for example the quality of the spatial XFEL beam profile and timing jitter may reduce the resolution. 

\begin{figure*}
\centering
  \includegraphics[width=\linewidth]{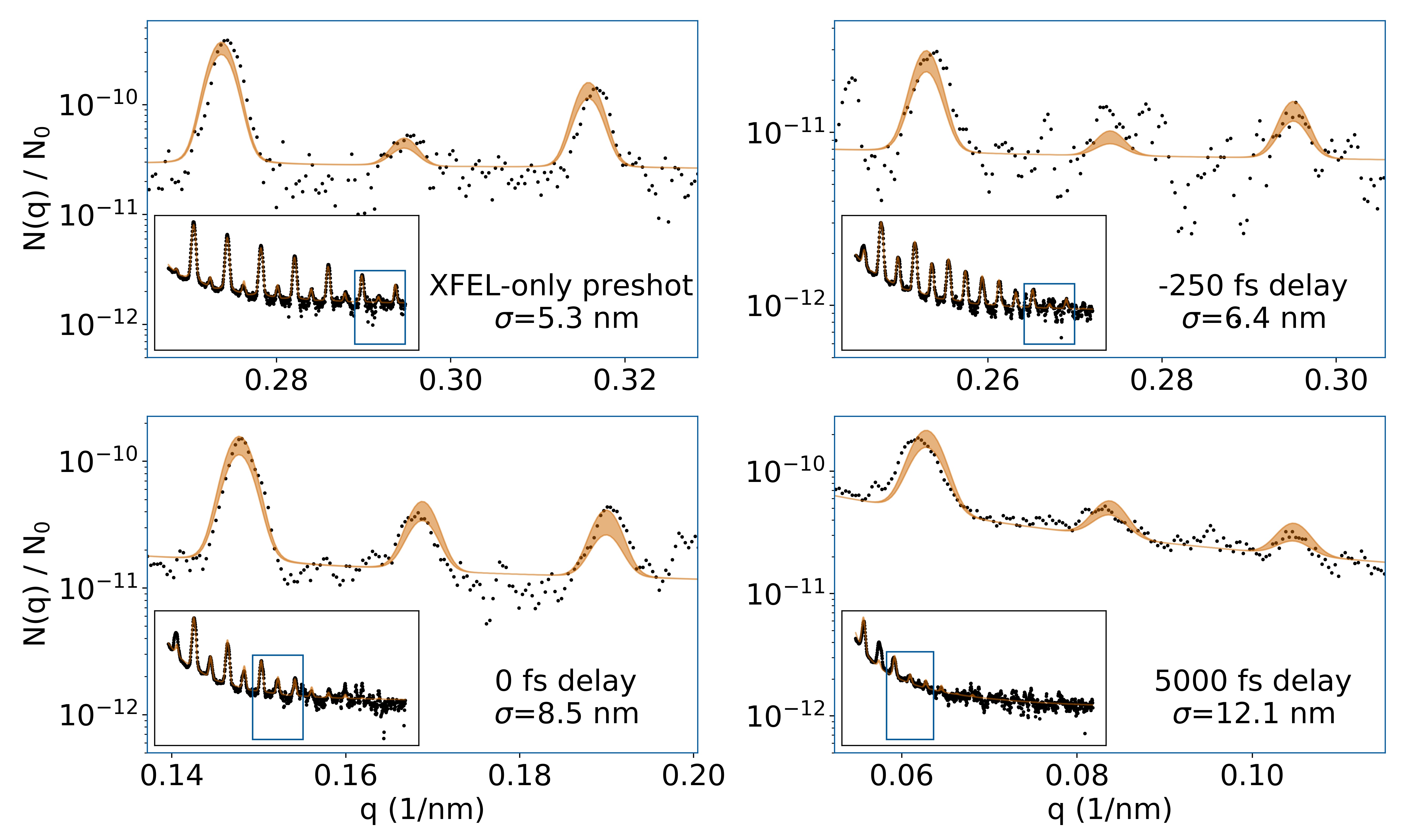}
\caption{Lineouts of scattered intensity normalized to the XFEL intensity (black points) taken along the $q$ direction perpendicular to the grating ridges, averaged over 20 pixels, the same as shown in Fig.~\ref{fig:results1}. The reminiscent background was modeled by the blue line and subtracted prior to the fit. 
The region shown enlarge the region marked in the respective insets to make visible the shaded area for the relevant peaks at large q-values. They represent the range of possible fits with a $\chi<1.01\chi_{min}$.}
  \label{fig:fitting}
\end{figure*}

\textbf{Fitting procedure and nanometer resolution}
Since the gratings did not provide an exact aspect ratio ridge width $b$ to grating period $g$ of exactly $1/2$ due to fabrication limits, it cannot be assumed that every even peak is extinct and every odd one is completely unaffected. 
Thus we cannot simply fit the edge scattering function (FT of the error function) to the scattering peak values. 
Rather, we must fit the full grating scattering equation~(1). 
We fixed the number of illuminated grating ridges, $N$, to the respective calculated values in the XFEL FWHM spot -- neglecting the shape of the XFEL intensity radial profile -- since it has almost no effect on values of the fitted width $\sigma$, and the point spread function of the detector is slightly larger than the FT of the illumination spot. 
The error function model for the density distribution was chosen based on the simulations (see below). 
However, as can be shown, for many other choices of the ridge edge shape the fit does not change significantly. 
In the fitting algorithm we also take into account a finite detector point spread function, which broadens and hence reduces the peak signal (in Fig.~\ref{fig:scan} we scaled the envelopes accordingly) and completely washes out the grating scattering side maxima. 
With the present detector it would only be possible to sample the fine side-maxima of the grating function if the distance between tha target and detector was increased to approx. $5\unit{m}$, which is technically not feasible due to space limitations and would reduce the angular angle covered by the detector drastically. 
Since the large scattering angles carry the information of small spatial scales, and we are especially interested in few nanometer expansion, we adopted the setup shown in Fig.~\ref{fig:setup}. 
Before fitting we subtract the electronic background (using separately recorded dark fields) and model the remaining X-ray background (e.g. from slit scattering at the XFEL cleaning apertures) with a power law.  
The individual heights of the peaks vary according to their phase in the single slit scattering function $S(q)$, which can reduce the fit sensitivity for $\sigma$ especially for large $\sigma$ where only a few maxima remain visible. 
This beat generates a correlation between $\sigma$ and $b/g$. 
It is possible to find different combinations of $\sigma$ and $b/g$ that lead to fits consistent with the data; this effect is included in the error bars in Fig.~\ref{fig:scan}. 
The accuracy in determining $\sigma$ is given by the ability to measure the exponential decrease at large $q$-values. 
Thus we achieve the highest accuracy, below a nanometer, for the smallest expansions at early times when many maxima are still visible. 
We determine the fit error by varying all fit parameters around the best fit values and require $\chi^2$ to remain within $1\%$ of the minimum $\chi^2_{min}$ at the best fit, which corresponds to the requirement for the respective range of the variation of the estimator to cover the range of the experimental data variation. 
The projection of the resulting volume in parameter space onto the $\sigma$-axis gives the confidence interval for $\sigma$. 
This error level, as shown in Fig.~\ref{fig:fitting}, is a reasonable choice to describe the range of deviation between the data and the fit curve. \\
%
%
%
%
%
%
%
%
The accuracy of determining the expansion parameter $\sigma$ is demonstrated by comparison of the two XFEL-only preshots that we recorded before each optical laser shot, measuring the inherent smoothness due to the production process. 
On average, $\sigma$ inferred from the two preshots varies only by $0.4\unit{nm}$, which represents the total stochastic error and hence demonstrates the accuracy of the method within a chosen model for the ion density profile. \\
For clarity, the error bars shown in Fig.~\ref{fig:scan} represent half of the $95\%$ interval of the delay jitter distribution in horizontal direction, and half of the $1\%$ deviation interval around $\chi^2_{min}$ in vertical direction. 
For the light blue data point no fit error could be determined but we still show the best fit value for completeness. 

\textbf{Numerical simulations}
Particle-in-cell simulations were performed using PICLS\cite{picls}, a fully relativistic 2D code including field ionization and direct impact ionization\cite{Mishra2013}. The target was modeled as a planar Si foil with a 200\,nm period grating of 100\,nm height, with the ridges having an error function shaped density edge with $\sigma$ at the experimental average value of un-driven gratings, determined by the XFEL preshots to be 5.4\,nm. We set 48 ions per cell, resulting in 672 electrons per cell at full ionization. The simulation box was set to $1\lambda\times \lambda$ with 256 cells/$\lambda$ and 256 time steps per laser period and periodic boundaries along the target surface. 
The laser was modeled as a spatially plane wave with a 80\,fs long Gaussian temporal profile, peak normalized field amplitude $a_0=0.38$ coming in from the left. We start the simulation $150\unit{fs}$ prior the arrival of the laser maximum on the target surface and follow the plasma evolution up to 2\,ps later. At this time the plasma expansion was stagnating, the gradient staying constant within $\pm1\unit{nm}$ over more than a picosecond. The shape of the edges remains error-function-like throughout the simulation. To extract the plasma expansion, we ignore the electron jets at times $>270\unit{fs}$ and only project the density of the plasma bulk. 
The jets consistently occur in the simulations for a range of larger laser intensities (we performed additional simulations up to an order of magnitude higher intensity) and longer pulse durations (up factor of two), always leading to a density pattern a few tens on nanometers in front of the surface following the preinscribed grating structure, shifted transversely by $g/2$. 

\begin{acknowledgements}
Support by the Ion Beam Center, HZDR,
is gratefully acknowledged. 
The authors like to thank L. Bischoff (HZDR) for his support in performing FIB cuts and electron microscopy of the target samples. 
This work was partially supported by DOE Office of Science, Fusion Energy Science under FWP 100182. The experiments were performed at the Matter at Extreme Conditions (MEC) instrument of LCLS, supported by the DOE Office of Science, Fusion Energy Science under contract No. SF00515. 
This work has also been supported by HIBEF (www.hibef.eu) and partially by Horizon 2020 LASERLAB-EUROPE/ LEPP (Contract No. 654148) and by the German Federal Ministry of Education and Research (BMBF) under contract number 03Z1O511. MG and MZ acknowledge support from the European Cluster of Advanced Laser Light Sources (EUCALL) project which has received funding from the European Union’s Horizon 2020 research and innovation programme under grant agreement No 654220.
\end{acknowledgements}


%

%
%


\end{document}